\documentclass[aps,prl,reprint,longbibliography,nofootinbib]{revtex4-2}
\usepackage{amsmath,amssymb}
\usepackage{microtype}
\usepackage{xcolor}
\usepackage[colorlinks=true,allcolors=blue]{hyperref}
\newcommand{\dd}{\mathrm{d}}
\newcommand{\E}{\mathcal E}

\begin{document}
\title{Nonzero cubic Love numbers of Schwarzschild black holes}
\author{Bruno Bucciotti}
\affiliation{Department of Physics and Beyond: Center for Fundamental Concepts in Science,
Arizona State University, Tempe, AZ 85281, USA}
\date{\today}
\begin{abstract}
Static Love numbers of Schwarzschild black holes vanish at linear and quadratic order in General Relativity, while generic nonlinear tides remain less understood. We solve a specific non-axisymmetric octupolar tide at cubic order and, building on the symmetry of static General Relativity exploited by Parra-Martinez and Podo, match the result to a worldline effective field theory (EFT) using dimensional regularization. Surprisingly, we find a Love number that neither vanishes nor runs.
\end{abstract}
\maketitle
\setcounter{secnumdepth}{-1}

Tidal deformability is a fundamental property of a gravitating object: its response to an external field probes its internal structure and determines how that structure affects its gravitational interactions. Tidal Love numbers quantify this response through the induced multipole moments. Four-dimensional black holes are exceptional because their static Love numbers vanish at linear order, despite the tidal distortion of their horizons. Evidence for the asymptotically flat, four-dimensional vanishing reaches back several decades: Damour explicitly obtained a vanishing quadrupolar Love number in 1983~\cite{Damour:1983compact}, noting agreement with D'Eath's 1975 calculation~\cite{DEath:1975background}. It was subsequently reexamined by Fang and Lovelace in 2005~\cite{Fang:2005qq}, established systematically in 2009~\cite{Binnington:2009bb,Damour:2009vw}, and confirmed by effective field theory (EFT) analyses~\cite{Kol:2011vg}. This vanishing is all the more puzzling because static black hole Love numbers are generically nonzero in higher-dimensional GR~\cite{Kol:2011vg,Hui:2020xxx}, in higher-curvature extensions of GR~\cite{Cardoso:2018eft,Cano:2025love}, and for asymptotically de Sitter and anti-de Sitter black holes~\cite{Nair:2024ds,Franzin:2025ads}. The study of black hole Love numbers has flourished in recent years, with advances in static responses and symmetry arguments~\cite{Hui:2020xxx,LeTiec:2021kerr,Chia:2021tidal,Charalambous:2021kerr,Porto:2016naturalness,Penna:2018carroll,Charalambous:2021kcz,Hui:2022ladder,BenAchour:2022hidden,Hui:2022nearzone,Charalambous:2022love,Berens:2023ladder,Katagiri:2023conformal,Solomon:2023duality,Sharma:2024ladder,Charalambous:2023myers,Charalambous:2024higher,Charalambous:2025branes,Lupsasca:2025love,Berens:2025geometric,Berens:2025higher,Kumar:2026highest}, scattering methods and dynamical matching, including scalar probes~\cite{Saketh:2022scattering,Ivanov:2023amplitudes,Ivanov:2023matching,Saketh:2024dynamical,Ivanov:2024raman,Correia:2025born,CaronHuot:2025born,Ivanov:2025tails,Correia:2025analyticity,Combaluzier:2025dynamical,Ivanov:2026toolkit}, and nonlinear gravitational response~\cite{Poisson:2010geometry,Gurlebeck:2015xpa,Poisson:2021operational,Poisson:2021allpn,Bern:2021nonlinear,Riva:2023rcm,Iteanu:2024dvx,Combaluzier-Szteinsznaider:2024sgb,Kehagias:2024rtz,Gounis:2024hcm,Parra-Martinez:2025bcu}, culminating in the comprehensive review by Rodr\'iguez, Santoni, and Solomon~\cite{Rodriguez:2026review}. Whether this absence of static tidal deformability survives nonlinear perturbations is therefore a fundamental question about black holes. In worldline EFT, Love numbers are defined as coefficients of local tidal operators describing a compact object represented by a point particle worldline~\cite{Goldberger:2004jt}, and are determined by matching the EFT field to the full gravitational solution with the same applied tide. Their vanishing has been phrased as a fine-tuning problem by Hui et al.~\cite{Hui:2020xxx}, motivating the search for an underlying symmetry. Hidden symmetries explain the vanishing at linear order~\cite{Charalambous:2021kcz}, while Riva et al. found no quadratic quadrupolar response~\cite{Riva:2023rcm} and Iteanu et al. extended the result to arbitrary multipoles at quadratic order~\cite{Iteanu:2024dvx}. Static axisymmetric solutions give a complementary constraint valid at every order~\cite{Gurlebeck:2015xpa,Poisson:2021allpn,Combaluzier-Szteinsznaider:2024sgb,Kehagias:2024rtz}, although they are not sufficient to determine every coupling in the worldline EFT. The EFT matching of Combaluzier--Szteinsznaider et al. identifies the combinations of nonlinear operators accessible to such tides~\cite{Combaluzier-Szteinsznaider:2024sgb}. Lastly, Parra-Martinez and Podo identified a spurionic symmetry, the Buchdahl inversion combined with a sign flip of the mass and of the applied tide. They used it to argue, in a unified way, that static Love numbers vanish and do not run at all orders, reproducing the known linear and quadratic results and the pattern of vanishing and running in higher dimensions~\cite{Parra-Martinez:2025bcu}.

In this Letter we compute the static response at cubic order for a non-axisymmetric electric\footnote{Here ``electric'' denotes a gravitational tide in the even-parity (polar or Zerilli) sector, described by the electric part of the Weyl tensor and its spatial derivatives, and does not refer to an electromagnetic field.} octupole, proceeding in two steps. First, we compute the solution regular at the horizon at cubic order. Second, we match this solution to the dimensionally regulated worldline EFT. In contrast with the all-order vanishing argued in~\cite{Parra-Martinez:2025bcu}, we obtain a nonzero Love coupling, which is finite and does not run. To locate the origin of this discrepancy, we examine critically each step in the application of their symmetry. We confirm that the field of the minimal point mass preserves it in dimensional regularization, also after minimal subtraction of the divergences and the change from harmonic to Regge--Wheeler gauge. However, we find that the solution regular at the horizon breaks the symmetry, contrary to what is assumed in~\cite{Parra-Martinez:2025bcu}, and this breaking is the whole reason why the Love number we compute is nonzero. To our knowledge, this is the first explicit cubic solution regular at the horizon for a non-axisymmetric tide, and the first to be matched to the worldline EFT.

An octupole has $\ell=3$, and the symmetric product of three octupoles contains two independent octupolar representations,
\begin{equation}
(\mathbf{3}\otimes\mathbf{3}\otimes\mathbf{3})_{\rm sym}
= 2\,\mathbf{3} \, \oplus \dots \, ,
\end{equation}
corresponding to the two worldline operators we will find. The bold labels denote angular momentum rather than representation dimension, and the subscript restricts the product to tensors symmetric under permutations of the three applied octupolar fields. There are consequently two independent worldline operators contributing to the cubic octupolar response, of which axisymmetric data constrain only one combination. Our non-axisymmetric calculation determines the remaining combination, allowing us to reconstruct the response to general octupolar data. This is the simplest case to treat, as a single unknown remains.

The paper is organized as follows: we describe the radial problem, introduce the static symmetry used by Parra-Martinez and Podo in our variables, and compute the solution regular at the horizon in General Relativity. We then match it to the worldline EFT, with explicit radial functions, coefficients of the reference field, and operator normalizations collected in the End Matter.

\section{Setup and homogeneous solutions}
We work at cubic order in the physical octupolar amplitude $\E$, keeping the exact mass dependence of the radial equations. Our units and radial variables are
\begin{equation}
G=1,\qquad x=\frac rM-1,\qquad f=1-\frac{2M}{r} \, .
\end{equation}
In Regge--Wheeler (RW) gauge the static metric is
\begin{equation}
\label{eq:metric}
\begin{split}
\dd s^2={}&-f(1+A)\dd t^2+f^{-1}(1+B)\dd r^2\\
&+2g_{ra}\dd r\dd x^a+r^2(1+K)\dd\Omega^2,
\end{split}
\end{equation}
where the angular index is $a$, and $g_{ra}$ is purely axial. We expand $A,B,K,g_{ra}$ as, for example,
\begin{equation}
A(r,\theta,\varphi)=A^{(1)}+A^{(2)}+A^{(3)}+O(\E^4) \, .
\end{equation}
The superscript $(n)$ includes the full amplitude dependence at order $n$, which we factor out while decomposing the angular dependence into spherical harmonics,
\begin{equation}
A^{(n)}=(\E M^3)^n\sum_{\ell,m}a_{\ell m}^{(n)}(x)Y_{\ell m}(\theta,\varphi) \, .
\end{equation}
We use the same decomposition for $B$ and $K$. After angular projection and elimination of the constraints, we solve at each order for the dimensionless polar radial field $a_{\ell m}^{(n)}(x)$,
\begin{equation}
\label{eq:radial}
\begin{split}
\mathcal L_\ell a_{\ell m}^{(n)}\equiv{}&(x^2-1)(a_{\ell m}^{(n)})''+2x(a_{\ell m}^{(n)})'\\
&-\left[\ell(\ell+1)+\frac4{x^2-1}\right]a_{\ell m}^{(n)}
=S_{\ell m}^{(n)} \, .
\end{split}
\end{equation}
Primes denote $x$ derivatives, and the source $S_{\ell m}^{(n)}$ vanishes at linear order, while at quadratic and cubic order it is fixed by the fields at lower orders. The operator $\mathcal L_\ell$ is invariant under $x\mapsto-x$, so its homogeneous solutions can be chosen even or odd. We impose horizon regularity and specify the growing amplitudes in the overlap region
\begin{equation}
|\E|M^3\ll1,\qquad M\ll r\ll|\E|^{-1/3} \, .
\end{equation}
Here and below ``at infinity'' denotes this formal expansion at large radius. For the octupole, we denote the $A$ components of the homogeneous tidal and decaying modes by $T_3$ and $D_3$, respectively. These radial functions are independent of $m$ and $n$, and obey
\begin{equation}
T_3\sim r^3,\quad D_3\sim r^{-4},\qquad r\gg M \, .
\end{equation}
This component convention applies throughout, and we write $K_3$ for the corresponding $K$ component of the octupolar tidal mode. The solution regular at the horizon and growing at infinity is
\begin{equation}
\label{eq:T}
T_3=x(x^2-1),\qquad
K_3=-T_3-x^2+\frac15 \, .
\end{equation}
We normalize the decaying solution by its $A$ component,
\begin{equation}
\label{eq:D}
\begin{split}
D_3={}&\frac12T_3\ln\frac{x+1}{x-1}-x^2+\frac23
+\frac{2}{15(x^2-1)}\\
={}&\frac8{105}\left(\frac Mr\right)^4
\left[1+\frac{4M}{r}+O\!\left(\frac{M^2}{r^2}\right)\right] \, .
\end{split}
\end{equation}
Under radial reflection $x\mapsto-x$, equivalently $r\mapsto2M-r$ at fixed $M$, $T_3$ is odd and $D_3$ is even.
Both modes are purely polar, so their axial components vanish, while the constraints fix their $B$ and $K$ components. The tidal mode is regular at the horizon, whereas $D_3$ is singular there.

We choose the applied tide to be the complex octupole
\begin{equation}
\label{eq:data}
\begin{gathered}
A^{(1)}=\E M^3T_3Y_{31},\quad B^{(1)}=-A^{(1)},\\
K^{(1)}=\E M^3K_3Y_{31},\\
Y_{31}=\sin\theta(5\cos^2\theta-1)e^{i\varphi} \, .
\end{gathered}
\end{equation}
Since the applied tide carries $e^{i\varphi}$ ($m=1$), its quadratic and cubic products carry $e^{2i\varphi}$ and $e^{3i\varphi}$, fixing $m=2$ and $m=3$ at these orders. For a real tide proportional to $\operatorname{Re}Y_{31}=\frac{Y_{31}+\bar Y_{31}}{2}$, the cubic $m=3$ result is one eighth of our complex one.

\section{The symmetry and the cubic solution}
Let us now describe the symmetry of the static equations identified in~\cite{Parra-Martinez:2025bcu}, a Buchdahl inversion combined with a radial reflection and a sign flip of the applied tide for even $\ell$. It becomes transparent in the variables
\begin{equation}
\label{eq:static}
\begin{gathered}
\dd s^2=-e^{2U}\dd t^2+e^{-2U}h_{ij}\dd x^i\dd x^j,\\
U=\frac12\ln[f(1+A)],\qquad U^{(0)}=\frac12\ln f \, ,
\end{gathered}
\end{equation}
where we expanded $U=U^{(0)}+U^{(1)}+U^{(2)}+U^{(3)}+O(\E^4)$ as we did for $A$, with $U^{(0)}\equiv u^{(0)}$ while $u_{\ell m}^{(n)}$ are defined as
\begin{equation}
U^{(n)}=(\E M^3)^n\sum_{\ell,m}u_{\ell m}^{(n)}(x)Y_{\ell m}(\theta,\varphi),
\qquad n\geq1 \, .
\end{equation}
The vacuum equations
\begin{equation}
\label{eq:vacuum}
\partial_i(\sqrt h\,h^{ij}\partial_jU)=0,\qquad
R_{ij}[h]=2\partial_iU\partial_jU
\end{equation}
are invariant under Buchdahl inversion, i.e. $U\mapsto-U$ at fixed $h_{ij}$~\cite{Buchdahl:1959nk}. Combining this inversion with the radial reflection $x\mapsto-x$, acting on $h_{ij}$ by the coordinate pullback, also preserves the Schwarzschild background while exchanging the horizon and the singularity. The induced action on $A$ is generically nonlinear: at linear order
\begin{equation}
	A^{(1)}(x)\mapsto-A^{(1)}(-x) \, ,
\end{equation}
whereas higher orders have non-homogeneous transformations.

In contrast, $U(x)\mapsto-U(-x)$ is linear, so a tidal mode of angular momentum $\ell$ transforms with $\sigma_\ell=(-1)^{\ell+1}$, and its decaying partner has the opposite sign. Transforming the applied tidal field with the same sign gives the combined transformation
\begin{equation}
\label{eq:reflection}
(x,U,h,\E_\ell)\longmapsto(-x,-U,h,\sigma_\ell\E_\ell) \, .
\end{equation}
We reserve ``radial reflection'' for $x\mapsto-x$ alone, while we call a solution ``symmetric'' or ``covariant'' when it is invariant under eq.~\eqref{eq:reflection}. For such solutions, at tidal order $n$ the radial functions of $U$ must then have parity $-\sigma_\ell^n$ under radial reflection. For our octupole $\sigma_3=1$, so $\E$ is unchanged and the radial profiles of $U$ must be odd in $x$ at every order in the tide. The RW conditions are also invariant, since Buchdahl inversion only rescales $g_{ij}$ and radial reflection preserves them.

At second order, $(\mathbf3\otimes\mathbf3)_{\rm sym}=\mathbf0\oplus\mathbf2\oplus\mathbf4\oplus\mathbf6$ leaves the polar channels $\ell=2,4,6$ for $m=2$. We select the polynomial solution regular at the horizon, with quadratic external data respecting the same symmetry. At third order the polar channels are $\ell=3,5,7,9$, and the solution regular at the horizon has a logarithm $\ln(r)$ only for $\ell=3$.

Normalizing the $3,3$ spherical harmonic as $Y_{33}=\sin^3\theta\,e^{3i\varphi}$, the solution regular at the horizon is
\begin{equation}
\label{eq:a3schem}
a_{33}^{(3)}=cT_3\ln(x+1)+\text{rational part},\qquad
c=\frac{448}{1287} \, .
\end{equation}
The rational part, given explicitly in app.~\ref{app:solution}, has a simple pole only at the Schwarzschild singularity. Reconstructing $u_{33}^{(3)}$ using lower order data gives the particularly simple decomposition (see app.~\ref{app:solution})
\begin{equation}
\label{eq:split}
u_{33}^{(3)}=\underbrace{\frac c4T_3\ln(x^2-1)+\mathcal R(x)}_{\text{symmetric}}
+\underbrace{\frac c2D_3}_{\text{non-symmetric}} \, .
\end{equation}
Here $\mathcal R(x)$ is odd in $x$, and the labels refer to the symmetry in eq.~\eqref{eq:reflection}, which acts on $U$ as $U(x)\mapsto-U(-x)$. The two terms are separately singular at the horizon, but their sum is regular. The presence of the non-symmetric term is one of the central results of this Letter: the symmetry of the horizon boundary condition assumed in~\cite{Parra-Martinez:2025bcu} would forbid this non-symmetric piece, whereas its coefficient $\frac{c}{2}$ is nonzero.

There is a structural reason why the non-symmetric part is a homogeneous mode. The original cubic solution and its image under the symmetry in eq.~\eqref{eq:reflection} have the same covariant lower orders and hence the same cubic source, so their difference solves the homogeneous equations. The symmetry in eq.~\eqref{eq:reflection} permits this defect because it exchanges the horizon boundary condition with a condition at the singularity. Thus the horizon condition need not preserve the symmetry, and eq.~\eqref{eq:split} shows that it does not.

\section{Matching the response}
Working in $D$ spacetime dimensions (dim-reg) to regulate divergences, we describe the compact object by the worldline action~\cite{Goldberger:2004jt}
\begin{equation}
S_{\rm pp}+S_{\rm tidal},\qquad S_{\rm pp}=-M\int\dd\tau \, ,
\end{equation}
where $\tau$ is proper time and $S_{\rm tidal}$ contains local couplings to the external curvature. The minimal action $S_{\rm pp}$ describes the reference point mass, while the Love couplings in $S_{\rm tidal}$ account for the difference between the field sourced by a minimal point particle and the full response of the object.

We define the electric octupole covariantly as $E_{ijk}=-\nabla_{\langle i}C_{j|\alpha|k\rangle\beta}v^\alpha v^\beta$ in an orthonormal rest frame of the worldline tangent $v^\mu$, with curvature sign chosen so that its limit in a weak field is
\begin{equation}
E_{ijk}=-\partial_{\langle i}\partial_j\partial_{k\rangle}U(0) \, ,
\end{equation}
where $U(0)$ only picks up the external tidal field, since mass-dependent terms vanish on the worldline in dim-reg (see app.~\ref{app:covariance}). Here $C_{\mu\nu\rho\sigma}$ is the Weyl tensor, and brackets mean symmetric trace-free (STF) projection. The quadratic and cubic parts of $S_{\rm tidal}$ vanish, as they encode linear and quadratic Love numbers~\cite{Iteanu:2024dvx,Combaluzier-Szteinsznaider:2024sgb}. As previously argued from group theory, the quartic part relevant to a pure electric octupole contains two independent operators, whose coefficients $k_a$ and $k_b$ are the cubic Love numbers. We normalize them as
\begin{equation}
\label{eq:operators}
\begin{split}
S_{\rm tidal}=\frac{M^{13}}{24}\int\dd\tau\,
\big[&k_a(E_{ijk}E_{ijk})^2\\
&+k_bE_{ijk}E_{ilm}E_{jln}E_{kmn}\big] \, .
\end{split}
\end{equation}

We now pass to isotropic coordinates, more suited for the computations that follow,
\begin{equation}
\label{eq:isotropic}
r=R\left(1+\frac M{2R}\right)^2,\qquad
x=\frac RM+\frac M{4R} \, ,
\end{equation}
In these coordinates, $M\mapsto-M$ at fixed $R$ coincides with $x\mapsto-x$ and behaves better in general dimensions $D$. Combining this mass sign change with Buchdahl inversion and the sign change of the applied tide gives the EFT counterpart of eq.~\eqref{eq:reflection}, which is the spurion symmetry of~\cite{Parra-Martinez:2025bcu}:
\begin{equation}
\label{eq:symmetry_eft}
(M,\E,U,h)\longmapsto(-M,-\E,-U,h) \, .
\end{equation}
For our octupole the product $\E M^3$ is unchanged, so eq.~\eqref{eq:symmetry_eft} acts on the radial profiles as $u(x)\mapsto-u(-x)$, exactly as eq.~\eqref{eq:reflection} does. As in the previous section, the background sourced by $S_{\rm pp}$ (see app.~\ref{app:covariance}) and the applied tide are covariant. What differs is the short-distance condition: horizon regularity is mapped to a condition at the singularity, whereas the EFT condition at $R=0$ is mapped to itself. Thus the field of the minimal point mass is symmetric, and the symmetry-breaking part of the EFT field must come entirely from $S_{\rm tidal}$.

We impose the spatial harmonic gauge $\partial_i(\sqrt h\,h^{ij})=0$, which is invariant under eq.~\eqref{eq:symmetry_eft}. For $\ell\geq2$ both harmonic and RW gauge fix the polar gauge ambiguity order by order, so the transformation to RW gauge preserves the symmetry.

Dimensional regularization also preserves the symmetry. A term in $U$ at tidal order $n$ and mass order $j$ scales as
\begin{equation}
\label{eq:masspowers}
\E^n(GM)^jR^{3n-j(D-3)} \, ,
\end{equation}
where the symmetry forces $n+j$ to be odd, the selection rule of~\cite{Parra-Martinez:2025bcu}, and we reintroduced Newton's constant $G$ to make the dimensions of $GM$ apparent. For angular momentum $\ell$, the homogeneous solutions for $U$ behave at large $R$ as
\begin{equation}
U_{\ell,\rm growing}\sim R^\ell,\qquad
U_{\ell,\rm decaying}\sim R^{-\ell-D+3} \, ,
\end{equation}
so at generic $D$ there is overlap with eq.~(\ref{eq:masspowers}) for $(n,j)=(1,0)$ and $(0,1)$. The first, $\E R^3$, changes the applied octupolar tide, while the second, $GM/R^{D-3}$, changes the monopole mass. These amplitudes are determined by keeping the linear tidal field and $M$ fixed. The remaining mass-dependent powers are generically distinct from homogeneous solutions, so their coefficients follow from the radial equations. The growing homogeneous solutions at higher tidal orders may still be added, with finite amplitudes specifying additional external boundary data. Once this data and the regulated point-mass prescription are fixed, the minimal theory has no independent response amplitude. In particular, the decaying octupole is fixed by localized multipolar sources, absent from $S_{\rm pp}$ and supplied by the tidal operators (app.~\ref{app:normalization}).

At $D=4$, additional coincidences between radial powers can generate logarithms, carrying information beyond the ordinary power series. For the cubic octupolar field, the possible growing and decaying logarithms have the forms
\begin{equation}
\begin{aligned}
(n,j)=(3,6):&\quad \E^3M^6R^3\ln(R/\mu),\\
(n,j)=(3,13):&\quad \frac{\E^3M^{13}}{R^4}\ln(R/\mu),
\end{aligned}
\qquad D=4 \, .
\end{equation}
The first describes a resonance with the applied tide, while the second would encode running of a cubic Love number.

In dimensional regularization, the growing resonance appears as a pole in $D-4$. The growing homogeneous solutions at higher tidal orders mentioned above can be added as counterterms to cancel this pole. Their finite parts specify the renormalized external amplitudes, whose scale dependence encodes the same information as the logarithm. In particular, the log coefficient is fixed by the radial source, while the finite amplitude at a chosen scale remains external boundary data. Since the symmetry holds for the complete regulated fields at generic $D$, it holds separately for every Laurent coefficient around $D=4$. Minimal subtraction therefore preserves it, including finite contributions from fields that vanish at $D=4$ multiplying poles. Tensor components that disappear at $D=4$ and counterterms induced at lower orders must all be retained before taking the limit, since vanishing factors multiplying poles can leave finite terms.

The decaying logarithm would instead require a response pole at $(n,j)=(3,13)$. This term is forbidden in the gravitational potential $U$ sourced by $S_{\rm pp}$ alone for every $D$, since $n+j=16$ is even, so its coefficient cannot contain a pole. Appendix~\ref{app:normalization} also shows that the quartic tidal insertion has no response pole: its source normalization is regular near $D=4$. Thus the cubic Love couplings have no running from either the minimal action or their own leading insertion, in line with the non-renormalization argument of~\cite{Parra-Martinez:2025bcu}.

We now match the part of the solution that breaks the symmetry, using the same applied field at linear order and the same growing homogeneous amplitudes at quadratic order in both descriptions. The cubic growing amplitude must cancel the pole described above, but its finite part need not be fixed for this matching. Changing that finite part adds a linear homogeneous tidal solution at order $\E^3$, with zero symmetry-breaking decaying response because the linear Love number vanishes.

We extract the EFT response in $U$ from eq.~\eqref{eq:tidalA} using $\Delta U^{(3)}=\Delta A^{(3)}/2$, since lower orders already agree in the common RW gauge. Its normalization follows from the leading $R^{-4}$ coefficient, which equals the $r^{-4}$ coefficient because $r=R+O(M)$. Comparing with the black hole $U$ in eq.~\eqref{eq:split} gives
\begin{equation}
\begin{aligned}
\left.u_{33,\rm BH}^{(3)}\right|_{\rm breaking}&=\frac c2D_3,\\
\left.u_{33,\rm EFT}^{(3)}\right|_{\rm breaking}&=\frac{3150k_b}{2}D_3 \, .
\end{aligned}
\end{equation}
Equating them fixes
\begin{equation}
\label{eq:response}
3150k_b = c = \frac{448}{1287} \, .
\end{equation}

For general octupolar data the operators induce the octupole moment
\begin{equation}
\label{eq:Q}
Q_{ijk}=M^{13}\big[k_a(E_{lmn}E_{lmn})E_{ijk}
+k_b\langle E_{ilm}E_{jln}E_{kmn}\rangle\big] \, .
\end{equation}
The $k_a$ term is proportional to the scalar $E_{lmn}E_{lmn}$, which vanishes for our complex tide, so our matching isolates $k_b$. To fix $k_a$ we invoke the vanishing of the EFT response for axisymmetric tides~\cite{Gurlebeck:2015xpa,Poisson:2021allpn,Combaluzier-Szteinsznaider:2024sgb,Kehagias:2024rtz}. Translating the axisymmetric octupole into our operator basis gives (see app.~\ref{app:normalization})
\begin{equation}
\label{eq:axisym}
50k_a+3k_b=0 \, .
\end{equation}
The two matching conditions fix both Love couplings. Combining them yields
\begin{equation}
\label{eq:result}
\boxed{\displaystyle k_a=-\frac{16}{2413125} \, ,\qquad
k_b=\frac{32}{289575} \, .}
\end{equation}
These couplings determine the response to real octupolar data through eq.~\eqref{eq:Q}, and are the main result of this paper.

\section{Discussion}
In this work we have determined the cubic octupolar response by computing $k_b$ directly and using the known axisymmetric constraint to fix $k_a$. The horizon-regular solution splits into a symmetric part, reproduced by the minimal point mass, and a decaying mode that breaks the symmetry and is matched to $S_{\rm tidal}$.

The nonzero cubic response originates from the horizon boundary condition, which breaks the symmetry in eq.~\eqref{eq:reflection}, whose EFT counterpart, eq.~\eqref{eq:symmetry_eft}, is preserved by the minimal worldline theory. Matching its breaking part separates the finite Love coupling from the logarithm associated with the applied tide. The absence of a response pole in dim-reg establishes that these static Love numbers do not run at the tidal and derivative orders considered here.

Our results are consistent with the vanishing of the linear~\cite{Binnington:2009bb,Damour:2009vw,Kol:2011vg} and quadratic~\cite{Riva:2023rcm,Iteanu:2024dvx} Love numbers, and with the absence of a response to axisymmetric tides at every order~\cite{Gurlebeck:2015xpa,Poisson:2021allpn,Combaluzier-Szteinsznaider:2024sgb,Kehagias:2024rtz}. Indeed, axisymmetric data probe only the combination of couplings in eq.~\eqref{eq:axisym}, which vanishes, and the nonzero Love number lives in the second combination, accessible only to non-axisymmetric tides.

Our analysis relies on the symmetry identified by Parra-Martinez and Podo~\cite{Parra-Martinez:2025bcu} and confirms its consequences for the worldline theory: the minimal point-mass field respects it, its selection rule applies and the cubic Love couplings do not run. However, we disagree with their conclusion that Love numbers vanish at all orders, and in fact our eq.~\eqref{eq:split} contradicts their argument that horizon regularity preserves the symmetry. The issue lies in the boundary condition at the horizon: eq.~\eqref{eq:reflection} maps regularity at the horizon to regularity at the singularity. The solution regular at the horizon is therefore symmetric only if it is also regular at the singularity, and at cubic order it is not, because of the logarithm.

The nonvanishing of the cubic Love number can also be understood without resorting to the explicit solution, provided the minimal point-particle EFT generates a $\ln(r/\mu)$ term at large distances. In full GR, the radial equations and their sources are built from the polynomial lower orders, and are thus rational in $r$ with poles only at $r=0$ and $2M$. Therefore the logarithmic branch cut starting from $r=\infty$ can only end at $0$ or $2M$. Horizon regularity then leaves only $\ln(r/\mu)$ exactly\footnote{In contrast with $\ln\big(\frac{r-2M}{\mu}\big)$, or some combination of the two logarithms.} and not only at large $r$, as confirmed by eq.~\eqref{eq:a3schem}. In particular, since $\ln(r/\mu)$ is neither even nor odd under the radial reflection $r\mapsto2M-r$ entering eq.~\eqref{eq:reflection}, the conclusion is that the full solution must contain terms incompatible with the minimal point-mass worldline.
Equation~\eqref{eq:split} gives the precise splitting.

Our results supply nonvanishing Wilson coefficients for post-Newtonian (PN) descriptions of compact binaries. The quartic octupolar operators $E_3^4$ we discussed only enter at 15PN order, but other quartic operators enter earlier.
One example is $(E_{ij}E_{ij})^2$, at 11PN. This operator is unique in four dimensions, and its coefficient vanishes by the axisymmetric results in~\cite{Gurlebeck:2015xpa,Poisson:2021allpn,Combaluzier-Szteinsznaider:2024sgb,Kehagias:2024rtz}. Operators mixing multipoles, schematically $E_2^2E_3^2$, enter from 13PN and are not all fixed by axisymmetric tides, but, because there are more than two of them, their derivation is more involved. Axial tides enlarge the set of operators further, again beyond what axisymmetric data constrain, and we leave their computation to future work.

This result opens the door to a promising and systematic study of the nonlinear black hole response, including additional cubic Love numbers in both the even and the odd sectors, and the search for a general closed formula. Other directions include sharpening our understanding of why the first two orders vanish, from the perspective of a symmetry broken at cubic order. Lastly, it would be interesting to compute quartic Love numbers and determine whether they run logarithmically or not.

\begin{acknowledgments}
\textit{Acknowledgments.---}The author thanks Lam Hui for comments on the manuscript and Julio Parra-Martinez for discussions at the workshop ``New Frontiers in Strong Gravity'', held in Benasque, 2026. The author is supported by the U.S. Department of Energy under grant number DESC0019470 and by the Heising--Simons Foundation ``Observational Signatures of Quantum Gravity'' collaboration grant 2024-5305. Claude and GPT were used to cross-check some computations and to polish the presentation. The author is solely responsible for the contents of this paper.
\end{acknowledgments}

\bibliography{biblio}

\clearpage
\raggedbottom
\setcounter{secnumdepth}{2}
\appendix
\numberwithin{equation}{section}
\renewcommand{\theequation}{\thesection\arabic{equation}}
\section*{End Matter}

\section{Radial source and rational solution}
\label{app:solution}
We set $y=\cos\theta$ and write the complete quadratic $U$ as
\begin{equation}
\begin{split}
U^{(2)}={}&(\E M^3)^2 (1-y^2) e^{2i\varphi}
\sum_{\ell=2,4,6}u_{\ell 2}^{(2)}(x)p_\ell(y),\\
(p_2,p_4,p_6)={}&\left(1,\frac52(7y^2-1),
\frac{35}{8}(33y^4-18y^2+1)\right),
\end{split}
\end{equation}
where the solution regular at the horizon and covariant under the symmetry in eq.~\eqref{eq:reflection} is
\begin{equation}
\begin{aligned}
u_{22}^{(2)}&=\frac{T_3}{105}(15x^2+19),\\
u_{42}^{(2)}&=\frac{4T_3}{1155}(55x^2-26),\\
u_{62}^{(2)}&=-\frac{T_3}{1155}(165x^2-49) \, .
\end{aligned}
\end{equation}
The other quadratic fields follow from the radial constraints. With these lower orders, the source in $\mathcal L_3a_{33}^{(3)}=S_{33}^{(3)}$ is
\begin{equation}
\label{eq:source}
\begin{split}
S_{33}^{(3)}=\frac8{6435}(x^2-1)\big(&7410x^7+42300x^6\\
+17460x^5-56310x^4&-16827x^3\\
+17800x^2&+4319x-934\big) \, .
\end{split}
\end{equation}
The complete rational part of eq.~\eqref{eq:a3schem} can be given compactly using the odd function already appearing in eq.~\eqref{eq:split}:
\begin{equation}
\label{eq:R}
\begin{split}
\mathcal R(x)=\frac{x}{19305}\big(&1380x^8+3990x^6-9848x^4\\
&-1826x^2+7648-\frac{448}{x^2-1}\big) \, .
\end{split}
\end{equation}
For the convention used here to fix the tidal mode,
\begin{equation}
\label{eq:aexplicit}
\begin{split}
a_{33}^{(3)}={}&cT_3\ln(x+1)+2\mathcal R(x)\\
&+c\left[-x^2+\frac23+\frac{2}{15(x^2-1)}\right]\\
&+\frac8{6435}T_3^2(-20x^3+705x^2+20x-467) \, .
\end{split}
\end{equation}
The last line is the projection of the products of fields at lower orders in
\begin{equation}
A^{(3)}=2U^{(3)}+A^{(1)}A^{(2)}-\frac13(A^{(1)})^3 \, .
\end{equation}
Although eq.~\eqref{eq:aexplicit} is written with denominators at both radial singular points, their horizon poles cancel.
%

\section{Point-like source and the symmetry}
\label{app:covariance}
For a worldline at rest at the origin, the static metric is
\begin{equation}
	\dd s^2=-e^{2U}\dd t^2+e^{-2U/(D-3)}h_{ij}\dd x^i\dd x^j \, ,
\end{equation}
and the equations of motion of $S_{\rm GR}+S_{\rm pp}$ are
\begin{equation}
\label{eq:EFT_EOM}
\begin{gathered}
\partial_i(\sqrt h\,h^{ij}\partial_jU)
=8\pi GM\frac{D-3}{D-2}e^{U(0)}\delta^{(D-1)}(x),\\
R_{ij}[h]=\frac{D-2}{D-3}\partial_iU\partial_jU \, .
\end{gathered}
\end{equation}
Evaluating $U(0)$ order by order in the mass and tide, radial terms scale schematically as 
\begin{equation}
U^{(n)}\sim\E^n(GM)^j R^{3n-j(D-3)} \, .
\end{equation}
At $R=0$ for $D<3$ these terms analytically continue to zero when $j\ge1$, while for $j=0$ we are looking at a tidal field with no black hole, so that the field is simply $\E R^3$ plus nonlinear corrections, all vanishing at $R=0$.
The resulting absence of nonlinear point-particle vertex contributions is the observation in footnote~2 of Ref.~\cite{Parra-Martinez:2025bcu}.
Thus $U(0)=0$ and $e^{U(0)}=1$ in the formal perturbative expansion. We highlight that this prescription, while correct, does not coincide with the full resummed result, which shows a singularity already at the horizon
\begin{equation}
\lim_{R\to \frac{M^+}{2}}U^{(0)}
=\lim_{R\to\frac{M^+}{2}}\ln\frac{1-M/(2R)}{1+M/(2R)}=-\infty \, .
\end{equation}
The regulated source for $U$ in eq.~(\ref{eq:EFT_EOM}) is therefore proportional to $M$ alone, so it changes sign with the left-hand side, while the equation for $h$ remains unchanged. Unlike horizon regularity, the ultraviolet prescription at $R=0$ is preserved by $M\mapsto-M$. With external data covariant under eq.~\eqref{eq:symmetry_eft} and $S_{\rm tidal}=0$, the perturbative point-mass solution is therefore covariant under the symmetry. Appendix~\ref{app:normalization} shows how the quartic tidal operators introduce a symmetry-breaking multipolar source at the worldline, equivalent to a symmetry-breaking ultraviolet boundary condition.

\section{Normalization of the tidal operators}
\label{app:normalization}
We vary the complete action $S_{\rm GR}+S_{\rm pp}+S_{\rm tidal}$. We define $P_{ijk}=\partial\mathcal L_{\rm tidal}/\partial E_{ijk}$, where $\mathcal L_{\rm tidal}$ is in eq.~\eqref{eq:operators}. At leading order in the weak-field expansion,
\begin{equation}
\begin{split}
\nabla^2U=8\pi G\frac{D-3}{D-2}\big[&M\delta^{(D-1)}(x)\\
&-P_{ijk}\partial_i\partial_j\partial_k\delta^{(D-1)}(x)\big] \, .
\end{split}
\end{equation}
Under the symmetry in eq.~\eqref{eq:symmetry_eft}, $P_{ijk}$ is even: it is cubic in $E_{ijk}$, which is odd like $U$, and proportional to the factor $M^{13}$ in eq.~\eqref{eq:operators}, which is also odd. The resulting tidal source is therefore even, whereas the equation for $U$ requires an odd source. This means that the terms in $S_{\rm tidal}$ break the symmetry as required to account for the second term in eq.~(\ref{eq:split}), in agreement with the spurion analysis of~\cite{Parra-Martinez:2025bcu}.

We now compute the coefficient of the decaying solution, sourced by $S_{\rm tidal}$. We choose $H_3=t_{ijk}x^ix^jx^k$ to be harmonic, with constant STF $t_{ijk}$. Integrating the distributional equation over a ball $B_\Lambda$ centered at the origin with radius $\Lambda\gg M$ gives
\begin{equation}
\begin{aligned}
48\pi G\frac{D-3}{D-2}t_{ijk}^*P_{ijk}
&=\int_{B_\Lambda}H_3^*\nabla^2U\,\dd^{D-1}x\\
&=\int_{B_\Lambda}\big[H_3^*\nabla^2U
-U\overbrace{\nabla^2H_3^*}^{0}\big]\,\dd^{D-1}x\\
&=\int_{B_\Lambda}\nabla\!\cdot\!\big(H_3^*\nabla U-U\nabla H_3^*\big)\,\dd^{D-1}x\\
&=\oint_{S_\Lambda}(H_3^*\partial_rU-U\partial_rH_3^*)\,\dd S \, .
\end{aligned}
\end{equation}

The monopole does not contribute because $H_3(0)=0$. We evaluate the surface integral using the leading decaying field and its radial derivative at large $r$. Setting $D=4$ gives the source normalization $24\pi t_{ijk}^*P_{ijk}$.

Writing $n^i=x^i/r$ for the unit radial vector, we obtain the induced field in terms of $Q_{ijk}$ defined in eq.~\eqref{eq:Q}:
\begin{equation}
U_{\rm response}=-\frac52\frac{Q_{ijk}n^in^jn^k}{r^4} \, .
\end{equation}
Here $Q_{ijk}=6\partial\mathcal L_{\rm tidal}/\partial E_{ijk}$ reproduces eq.~\eqref{eq:Q} when the derivative is taken over the independent STF components. On our complex data,
\begin{equation}
E_{ijk}n^in^jn^k=-3\E Y_{31},\qquad
E_{ijk}E_{ijk}=0 \, ,
\end{equation}
\begin{equation}
\langle E_{ilm}E_{jln}E_{kmn}\rangle n^in^jn^k
=-48\E^3Y_{33} \, .
\end{equation}
At leading weak-field order the induced spatial metric is $\delta g_{ij}=-2U_{\rm response}\delta_{ij}$, which already satisfies the polar RW conditions. Conversion at higher mass orders therefore cannot change the leading $r^{-4}$ coefficient.  Away from the worldline the induced metric perturbation obeys the homogeneous linearized Einstein equations on Schwarzschild, so its decaying $A$ component in RW gauge is proportional to $D_3$.

We denote the difference between the EFT fields with and without the quartic tidal operators by $\Delta U^{(3)}$ and $\Delta A^{(3)}$, in RW gauge. Eq.~\eqref{eq:static} gives $\Delta A^{(3)}=2\Delta U^{(3)}$ exactly at cubic order, and its leading weak-field limit is $\Delta U^{(3)}=U_{\rm response}$, so comparison with eq.~\eqref{eq:D} fixes
\begin{equation}
\label{eq:tidalA}
\Delta A^{(3)}=3150k_b(\E M^3)^3D_3Y_{33} \, .
\end{equation}
The breaking response vanishes up to quadratic order, so eq.~\eqref{eq:tidalA} is the complete expression at cubic order. The explicit expression also clarifies that there is no pole at $D=4$, which, together with the absence of a response pole in the minimal point-mass field, establishes that the cubic Love number is finite and does not run.

Lastly, we discuss the constraint coming from the axisymmetric case. For an axisymmetric octupolar tide we write
\begin{equation}
E_{ijk}=e\,T_{ijk},\qquad T_{ijk}n^in^jn^k=P_3(\cos\theta) \, ,
\end{equation}
where $T_{ijk}$ is a constant STF tensor and $e$ is a real amplitude, so that the potential is $U=-\frac e6r^3P_3(\cos\theta)$ at leading weak-field order. The value of $e$ drops out of the constraint below, which is homogeneous in it. The nonzero components of $T_{ijk}$ are $T_{zzz}=1$ and $T_{zxx}=T_{zyy}=-1/2$, including permutations. Hence $T_{ijk}T_{ijk}=1+6/4=5/2$. We define $V_{ijk}=T_{ilm}T_{jln}T_{kmn}$. Direct contraction gives $V_{zzz}=3/4$ and $V_{iiz}=1$, so its STF projection has $V_{\langle zzz\rangle}=3/4-3/5=3/20$. Axisymmetry leaves a unique rank-three STF tensor, implying $V_{\langle ijk\rangle}=(3/20)T_{ijk}$. Equation~\eqref{eq:Q} therefore gives
\begin{equation}
Q_{ijk}=M^{13}e^3\Big(\frac52k_a+\frac3{20}k_b\Big)T_{ijk} \, .
\end{equation}
The vanishing axisymmetric response~\cite{Gurlebeck:2015xpa,Poisson:2021allpn,Combaluzier-Szteinsznaider:2024sgb,Kehagias:2024rtz} thus imposes in our normalization
\begin{equation}
\frac52k_a+\frac3{20}k_b=0 \, .
\end{equation}
Together with $3150k_b=c$, this fixes eq.~\eqref{eq:result}.
\end{document}